\documentclass[journal]{IEEEtran}

\usepackage{amsmath}
\usepackage{amssymb}
\usepackage{amsfonts}
\usepackage{graphicx}
\usepackage{epsfig}
\usepackage{subfigure}
\usepackage{psfrag}
\usepackage{cite}
\usepackage{latexsym}
\usepackage{url}
\usepackage{color}
\usepackage{multirow}
\usepackage{mathtools}
\usepackage{bm}
\usepackage{booktabs}
\usepackage{algorithm}
\usepackage{algpseudocode}
\usepackage{bm}

\usepackage{verbatim}
\usepackage{indentfirst}
\usepackage{hyperref}
\usepackage{multirow}
\usepackage{bbm}

\graphicspath{{fig/}}

\PassOptionsToPackage{bookmarks={false}}{hyperref}

\IEEEoverridecommandlockouts
\newcommand{\mv}[1]{\mbox{\boldmath{$ #1 $}}}

\allowdisplaybreaks[4] 

\begin{document}
\title{Pinching-Antenna Systems (PASS) Aided Over-the-air Computation}
\author{
Zhonghao Lyu, Haoyun Li, Yulan Gao, Ming Xiao, \emph{Senior Member, IEEE}, and H. Vincent Poor, \emph{Life Fellow, IEEE} \\
\thanks{Z. Lyu, H. Li, Y. Gao, and M. Xiao are with the Department of Information Science and Engineering, KTH Royal Institute of Technology, Stockholm, Sweden (e-mail: lzhon@kth.se, yulang@kth.se, haoyunl@kth.se, mingx@kth.se). }
\thanks{H. V. Poor is with the Department of Electrical and Computer
Engineering, Princeton University, New Jersey 08544, USA (e-mail:
poor@princeton.edu).}
}
\maketitle
\begin{abstract}
Over-the-air computation (AirComp) enables fast data aggregation for edge intelligence applications. However the performance of AirComp can be severely degraded by channel misalignments.
Pinching antenna systems (PASS) have recently emerged as a promising solution for physically reshaping favorable wireless channels to reduce misalignments and thus AirComp errors, via low-cost, fully passive, and highly reconfigurable antenna deployment. Motivated by these benefits, we propose a novel PASS-aided AirComp system that introduces new design degrees of freedom through flexible pinching antenna (PA) placement. 
To improve performance, we consider a mean squared error (MSE) minimization problem by jointly optimizing the PA position, transmit power, and decoding vector. To solve this highly non-convex problem, we propose an alternating optimization based framework with Gauss-Seidel based PA position updates. Simulation results show that our proposed joint PA position and communication design significantly outperforms various benchmark schemes in AirComp accuracy.
\end{abstract}
\begin{IEEEkeywords}
Pinching antenna systems (PASS), over-the-air computation (AirComp), data aggregation, multiple-access channels (MAC).
\end{IEEEkeywords}

\section{Introduction}
The proliferation of intelligent Internet-of-Things (IoT) devices and edge artificial intelligence (AI) applications is accelerating the evolution of the sixth-generation (6G) wireless networks towards task-oriented information processing for ultra-low-latency services \cite{GzhuZLyu,JPark}. In this context, over-the-air computation (AirComp) has emerged as a promising solution to enable low-latency and bandwidth-efficient wireless data aggregation. By exploiting the waveform superposition property of wireless multiple-access channels (MACs), AirComp enables simultaneous analog transmission and  functional aggregation of distributed data, thus realizing the compute-when-communicate paradigm \cite{XCao2024}. This makes it highly suitable for real-time AI services such as federated fine-tuning for large models, environmental monitoring, and autonomous control.

Despite its benefits, the practical implementation of AirComp faces critical challenges. Specifically, the performance of AirComp relies heavily on precise signal alignment among distributed devices. To this end, channel mismatches caused by fading or asynchronization can significantly degrade the accuracy of target functions \cite{ZWang2024}. Therefore, enhancing the underlying wireless channels is  essential to enable robust and scalable AirComp.

To address this issue, recent efforts have explored reshaping wireless environments to support AirComp, including reconfigurable intelligent surfaces (RISs), amplify-and-forward (AF) relays, and movable antenna (MA)/fluid antenna (FA) systems. Specifically, in \cite{Wzhang2022,GChen2025}, RISs achieve passive beamforming gains by adjusting the phase shifts of reflective elements to suppress AirComp mean squared errors (MSEs). Similarly, relay-assisted systems enhance AirComp accuracy by compensating for poor direct channels \cite{ZLin2022}. Moreover, MA systems have been investigated in \cite{NLi2025} for their potential to improve channel quality by locally displacing antennas, thereby forming more desirable channels for data aggregation in AirComp.

Recently, a new antenna technology, pinching antenna systems (PASS), has been proposed to physically reshape wireless channels. PASS utilizes simple dielectric pinchers clamped onto a dielectric substrate, thus forming a reconfigurable antenna array, where pinching antenna (PA) positions can be flexibly adjusted to manipulate the array response and induce spatial diversity \cite{ZYang,YLiu2025}. PASS has several unique advantages: First, it provides high reconfigurability with ultra-low deployment cost, as the position of  PAs can be effortlessly adjusted without any electrical reconnection. Second, PASS operates in a fully passive manner, eliminating the need for dedicated power or control circuits. Third, PASS enhances line-of-sight (LoS) propagation and supports a novel form of position-based beamforming, which enables precise spatial signal control. These benefits have been successfully explored in multi-user \cite{JZhao}, multiple-input-multiple-output (MIMO)  \cite{ABereyhi}, non-orthogonal multiple access (NOMA) \cite{SHu}, and secure communication systems \cite{GZhu}.

Given these properties, PASS is an attractive candidate for enhancing AirComp by physically shaping effective wireless channels to reduce misalignment and thus computation errors, while maintaining system simplicity and scalability. In particular, the spatial flexibility of PASS introduces new design degrees of freedom (DoFs) for AirComp. 
However, PASS-aided AirComp is in its infancy. One major challenge lies in the joint optimization of PA positions and communication resources, which is essential for fully unlocking the potential of deploying PASS in AirComp. This task is highly non-trivial due to the nonlinear dependence of the channel coefficients on PA positions, which simultaneously influence both the amplitude and phase. Furthermore, the computation error minimization problem in PASS-aided AirComp is inherently highly non-convex, due to the close coupling between PA position and communication variables in the co-design framework.

We propose a PASS-aided AirComp system, where a PASS-enabled base station (BS) serves as the fusion center computing functions over simultaneously transmitted user data. We aim to minimize the AirComp MSE through the joint optimization of PA positions, transmit power, and decoding vector, which is highly non-convex and thus non-trivial to be optimally solved. To tackle it, we propose an efficient alternating optimization (AO) framework with Gauss-Seidel-based PA position adjustment. Numerical results demonstrate that the proposed design significantly outperforms benchmark schemes in AirComp accuracy, highlighting the potential of PASS to enable scalable, reconfigurable, and cost-effective AirComp systems.

\section{System Model and Problem Formulation}\label{Sec-SM}

\begin{figure}[h]
	\centering
	 \epsfxsize=1\linewidth
		\includegraphics[width=8cm]{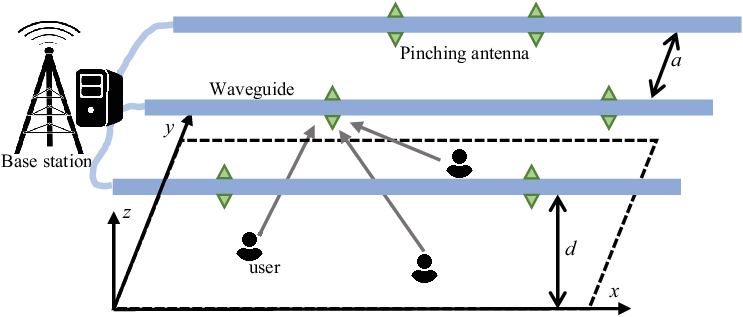}
		\vspace{-7pt}
	\caption{\label{model} Schematic of PASS-aided AirComp.}
\end{figure}

We consider a PASS-aided AirComp system operating over an MAC. Specifically, a BS is equipped with $M$ parallel dielectric waveguides, each equipped with $N$ PAs, and aims to aggregate information from a set of $K \geq 1$ users, denoted by $\mathcal{K} \triangleq \{1, \dots, K\}$, where each device is equipped with a single antenna. Consider a three-dimensional (3D) coordination system, where the users are uniformly distributed over a rectangular area on the $x$-$y$ plane with sides $L_x$ and $L_y$, and denote the position of the $k$-th user as $\mv{u}_k = (x_k, y_k, 0)$. Without loss of generality, we assume the waveguides in the PASS are deployed along the $x$-axis  at height $d$, and each with a length of $L_x$. Moreover, we assume the spacing between adjacent waveguides is $a$. Then the location of the $n$-th PA on the $m$-th waveguide is given by $\mv{v}_{m,n} = (v_{m,n}, (m-1)a, d)$, where $0 \leq v_{m,n} \leq L_x$. The feed point of each waveguide is located at $\mv{v}_{m,0} = (0, (m-1)a, d)$.

The objective of the BS is to recover the summation of the transmitted data from all users by leveraging the waveform superposition property of the wireless MAC. Specifically, let $s_k$ denote the data symbol transmitted by device $k \in \mathcal{K}$. Without loss of generality, we assume that $\{s_k\}$ are independent circularly symmetric complex Gaussian (CSCG) random variables with zero mean and unit variance, i.e., $s_k \sim \mathcal{CN}(0,1), \forall k \in \mathcal{K}$. Then the objective of the BS is to compute 
\begin{align}\label{target}
    s =  \sum\nolimits_{k=1}^K s_k.
\end{align}

To fulfill the above objective, we proceed to model the uplink transmission process.  Denote $h_{k,m,n}$ as the channel coefficient between user $k$ and the $n$-th pinching element on waveguide $m$, i.e., 
\begin{align}
	h_{k,m,n}=\frac{\lambda e^{-j \frac{2 \pi}{\lambda}\|{\mv u}_k-{\mv v}_{m,n}\|}}{4\pi \|{\mv u}_k-{\mv v}_{m,n}\|},
\end{align}
where the wavelength is given by $\lambda=c/f$, with 
$f$ and $c$ representing the carrier frequency and the speed of light, respectively.
Then, the received signal at the $n$-th PA on waveguide $m$ is given as
\begin{align}
    r_{m,n} = \sum\nolimits_{k=1}^K h_{k,m,n} \sqrt{p_k}s_k + z_{m,n},
\end{align}
where $p_k$ is the transmit power of user $k$, $z_{m,n}$ is the additive Gaussian white noise (AWGN) with variance $\sigma^2$, i.e., $z_{m,n} \sim \mathcal{CN} (0, \sigma^2)$.
Following the assumption that the waveguides are fully passive, the aggregated signal received by all PAs on waveguide $m$ at its feed point ${\mv v}_{m,0}$ is
\begin{align}
    \tilde{r}_{m} &= \sum\nolimits_{n=1}^N r_{m,n} e^{-j \frac{2 \pi i_{ref}}{\lambda}\|{\mv v}_{m,n}-{\mv v}_{m,0}\|}  \\
    &= \sum_{k=1}^K   \!\sum_{n=1}^N \sqrt{p_k} h_{k,m,n} e^{-j \frac{2 \pi i_{ref}}{\lambda}\|{\mv v}_{m,n}-{\mv v}_{m,0}\|}s_k \!+ \tilde{z}_{m},
\end{align}
where $i_{\rm ref}$ is the refractive index of  waveguides, and $ \tilde{z}_{m} = \sum_{n} z_{m,n} \sim \mathcal{CN} (0, N\sigma^2)$ is the cumulative noise over all PAs on waveguide $m$. 
Denoting the PA positions on waveguide $m$ as $\tilde{\mv v}_{m}=[{ v}_{m,1}, \cdots, { v}_{m,N}]^{\rm T} \in \mathbb{R}^{N \times 1}$, the equivalent channel between user $k$ and the feed point of waveguide $m$ is
\begin{align}
&g_{m,k}(\tilde{\mv v}_{m}) \nonumber \\ &= \sum\nolimits_{n=1}^N  h_{k,m,n} e^{-j \frac{2 \pi i_{ref}}{\lambda}\|{\mv v}_{m,n}-{\mv v}_{m,0}\|}, \nonumber \\
&= \sum\nolimits_{n=1}^{N}   \frac{\lambda e^{-j \left(\frac{2 \pi}{\lambda}\|{\mv u}_k-{\mv v}_{m,n}\|+ \frac{2 \pi i_{ref}}{\lambda}\|{\mv v}_{m,n}-{\mv v}_{m,0}\|  \right)}}{4\pi \|{\mv u}_k-{\mv v}_{m,n}\|}, \nonumber \\
&  =   \!\!\sum_{n=1}^{N}  \! \!  \frac{\lambda e^{ \!-j \! \left(\frac{2 \pi}{\lambda} \!\sqrt{(v_{m,n} \!- \!x_k)^2+((m-1)a-y_k)^2+d^2} +  \!\frac{2 \pi i_{ref}}{\lambda}  \!v_{m,n}   \!\right)}}{4\pi \sqrt{(v_{m,n}-x_k)^2+((m-1)a-y_k)^2+d^2}}.
\end{align}
Then we re-write $\tilde{r}_{m}$ with $g_{m,k}(\tilde{\mv v}_{m})$ as
\begin{align}
\tilde{r}_{m} &= \sum_{k=1}^K \sqrt{p_k}g_{m,k} (\tilde{\mv v}_{m}) s_k + \tilde{z}_{m} = {\mv g}^{\rm T}_m(\tilde{\mv v}_{m}) \mv P {\mv s} + \tilde{z}_{m}, 
\end{align}
where ${\mv g}_m(\tilde{\mv v}_{m})=[g_{m,1}(\tilde{\mv v}_{m}),\cdots,g_{m,K}(\tilde{\mv v}_{m})]^{\rm T} \in \mathbb{C}^{K \times 1}$ is the equivalent channel vector from all users to the feed point of waveguide $m$, $\mv P= {\rm diag}[\sqrt{p_1},\cdots,\sqrt{p_K} ]$ is the diagonal transmit power matrix,  and $\mv{s}=[s_1,\cdots,s_K]^{\rm T} \in \mathbb{C}^{K \times 1}$. Then, we compactly represent the effective uplink channel as
\begin{align}
    \mv r = \mv{G} (\mv{V}) \mv P \mv s + \mv z,
\end{align}
where $\mv r=[\tilde{r}_{1}, \cdots, \tilde{r}_{M}]^{\rm T}\in \mathbb{C}^{M \times 1}$, $\mv V= [\tilde{\mv v}_{1}, \cdots, \tilde{\mv v}_{M}]^{\rm T} \in \mathbb{R}^{M \times N}$, $\mv{G} (\mv{V})=[{\mv g}_1(\tilde{\mv v}_1),\cdots,{\mv g}_M(\tilde{\mv v}_{M})]^{\rm T} \in \mathbb{C}^{M \times K}$, and $\mv z=[\tilde{z}_{1}, \cdots, \tilde{z}_{M}]^{\rm T} \in \mathbb{C}^{M \times 1}$. 

Upon receiving the signal $\mv r$, the BS applies a decoding vector $\mv w$ to recover the summation of the transmitted signal, 
\begin{align}\label{estimated}
\hat{s}= \mv{w}^{\rm H} \mv{G} (\mv{V}) \mv P \mv s +  \mv{w}^{\rm H} \mv z.
\end{align}
The distortion between the target in \eqref{target} and the estimated value in \eqref{estimated} is measured by the MSE, defined as
\begin{align}
    \mathbb{MSE}(\mv w, \mv V, \mv P) & \triangleq {\mathbb E}[|\hat s - s|^2] \nonumber \\
    & = {\mathbb E} \left[\left| (\mv{w}^{\rm H} \mv{G} (\mv{V}) \mv P -{\mv 1}^{\rm T})\mv s+ \mv{w}^{\rm H} \mv z \right|^2 \right] \nonumber \\
    & = \|\mv{w}^{\rm H} \mv{G} (\mv{V})\mv P -{\mv 1}^{\rm T}\|^2 + {N\sigma^2}\|\mv w\|^2,
\end{align}
where $\mv 1=[1, \cdots,1]^{\rm T} \in \mathbb{R}^{K \times 1}$ denotes the all-ones vector.

Our objective is to minimize the MSE by jointly optimizing the positions of PAs $\mv V$, as well as the transmit power matrix $\mv P$ and decoding vector $\mv w$, subject to transmit power and physical placement constraints on the PAs. The corresponding optimization problem is formulated as
\begin{subequations}\label{P1}
	\begin{align}
	\text{(P1)}: \mathop {\min }\limits_{\mv w, \mv P, \mv V} &   \|\mv{w}^{\rm H} \mv{G} (\mv{V})\mv P -{\mv 1}^{\rm T}\|^2 + {N\sigma^2}\|\mv w\|^2 \nonumber \\
	\mathrm{s.t.}~&  \mv P^2 \preceq \mv P^{\rm max} \label{P1-a}\\
	~
	& 0 \le v_{m,n} \le L_x, \forall m,n \label{P1-b}\\
	& v_{m,n} \!-\! v_{m,n-1} \!\ge\! L_0, \!\forall m, \!\forall n \!=\! 2, \cdots, N, \label{P1-c}
	\end{align}
	\end{subequations}
where \eqref{P1-a} is a per-user power constraint with $\mv{P}^{\rm max} = \mathrm{diag}[P^{\rm max}_1, \cdots, P^{\rm max}_K]$, and $P^{\rm max}_k$ denotes the maximum transmit power of user $k$, \eqref{P1-b} defines the feasible deployment region of the PAs, \eqref{P1-c} enforces a minimum spacing $L_0$ between adjacent PAs on each waveguide to prevent antenna coupling effects.

\section{Joint PA Position and Communication Design}

Problem (P1) is highly non-convex due to the complex structure of $G(\mv{V})$, where the PA positions influence both the amplitude and phase components of the channel coefficient. Additionally, the close coupling among the PA position matrix $\mv{V}$, transmit power matrix $\mv{P}$, and decoding vector $\mv{w}$ further complicate the optimization. To tackle this challenge, we adopt an AO framework, where the PA positions are efficiently updated using a Gauss-Seidel-based iterative strategy.

\subsubsection{Optimization of the decoding vector $\mv w$} We consider the optimization of the decoding vector $\mv w$  under given transmit power and PA positions $\mv P$ and $\mv V$, for
which the optimization problem becomes
\begin{subequations}\label{P2}
	\begin{align}
	\text{(P2)}: \mathop {\min }\limits_{\mv w} &  \|\mv{w}^{\rm H} \mv{G} (\mv{V})\mv P -{\mv 1}^{\rm T}\|^2 + {N\sigma^2}\|\mv w\|^2 \nonumber.
\end{align}
\end{subequations}

Problem (P2) is a least-squares problem and hence convex. By setting the derivative of the objective function with respect to (w.r.t.) $\mv w$ to $\mv 0$, i.e., 
\begin{align}
&\frac{ \partial \mathbb{MSE}(\bm w)}{\partial \mv w} \nonumber \\
& \!=\!\! {2} \!\left(  \mv{G} (\mv{V})\mv P {\mv P}^{\rm T}\mv{G}^{\rm H} (\mv{V}) \mv w \!-\! \mv{G} (\mv{V})\mv P \mv{1} + N\sigma^2 \mv w\right) \!=\! \mv 0,
\end{align}
we have the optimal $\mv {w}^{*}$ under fixed $\mv P$ and $\mv V$ as
\begin{align}\label{Optimal-m}
\mv w^{*} = \left( \mv{G} (\mv{V})\mv P {\mv P}^{\rm T}\mv{G}^{\rm H} (\mv{V})  + N \sigma^2 {\mv I}_M \right)^{-1} \mv{G} (\mv{V})\mv P \mv{1}.
\end{align}

\subsubsection{Optimization of the transmit power $\mv P$} We fix the PA positions $\mv L$ and the decoding vector  $\mv w$, and then optimize the transmit power matrix $\mv P$ as
\begin{subequations}\label{P3}
	\begin{align}
	\text{(P3)}: \mathop {\min }\limits_{\mv P} &  \|\mv{w}^{\rm H} \mv{G} (\mv{V})\mv P -{\mv 1}^{\rm T}\|^2  \nonumber \\
	\mathrm{s.t.}~&  (11 {\rm a}). \nonumber
	\end{align}
\end{subequations}

Problem (P3) is a quadratically constrained quadratic programming (QCQP) problem, which is convex and can be optimally solved using the Karush-Kuhn-Tucker (KKT) conditions. To this end, we first construct the Lagrangian function associated with problem (P3) as follows,
\begin{align}
\mathcal{L}(\mv P, \bm{\mathit{\Lambda}})& =
 \mv{w}^{\rm H} \mv{G}(\mv{V}) \mv P \mv P^{\rm T} \mv{G}^{\rm H}(\mv{V})\mv{w}
\nonumber \\
&\!-  \!\!2\mathrm{Re} \left\{ \mv{w}^{\rm H} \mv{G}(\mv{V}) \mv P \mathbf{1}\right\}
\!+\! \rm{tr}\left(\bm{\mathit{\Lambda}} (\mv P^2 \!-\! \mv P^{\rm max}) \right),
\end{align}
where $\bm{\Lambda} = \mathrm{diag}[\lambda_1, \cdots, \lambda_K]$ is a diagonal matrix consisting of Lagrange multipliers associated with the power constraints. The corresponding Karush-Kuhn-Tucker (KKT) conditions are given by
\begin{align}
\frac{ \partial \mathcal{L}(\mv P, \bm{\mathit{\Lambda}})}{\partial \mv P} &= 2  \mv{G}^{\rm H}(\mv{V})\mv{w} \mv{w}^{\rm H} \mv{G}(\mv{V}) \mv P- 2\mathrm{Re}(\mv{G}^{\rm H}(\mv{V})\mv{w}\mathbf{1}^{\mathrm{T}}) \nonumber \\
&+2 \bm{\mathit{\Lambda}} \mv P = \mv 0 \label{KKT-1}\\
& \bm{\mathit{\Lambda}} \succeq 0, \bm{\mathit{\Lambda}} (\mv P^2 - \mv P^{\rm max}) = \mv 0, \label{KKT-2}\\
&\mv P^2 \preceq \mv P^{\rm max}. \label{KKT-3}
\end{align}
Then, the optimal transmit power $\mv P^*$ from \eqref{KKT-1} is
\begin{align}\label{opt_power}
\mv P ^{*}=&\big({\rm diag}(\mv{G}^{\rm H}(\mv{V})\mv{w} \mv{w}^{\rm H} \mv{G}(\mv{V}))+\bm{\mathit{\Lambda}} \big)^{-1} \nonumber \\
&{\rm diag}({\rm Re}(\mv{G}^{\rm H}(\mv{V})\mv{w})),
\end{align}
where the matrix of Lagrange multipliers $\bm{\mathit{\Lambda}}$ should be chosen to satisfy \eqref{KKT-2} and \eqref{KKT-3}. By substituting the optimal transmit power matrix \eqref{opt_power} into \eqref{KKT-2}, we have
\begin{align}
\bm{\mathit{\Lambda}}^{*}= & \bigg(\!{\rm diag}({\rm Re}(\mv{G}^{\rm H}(\mv{V})\mv{w})) (\mv P^{\rm max})^{-\frac{1}{2}}\! \! \nonumber \\
&-\!{\rm diag}(\mv{G}^{\rm H}(\mv{V})\mv{w} \mv{w}^{\rm H} \mv{G}(\mv{V})) \bigg)_+,
\end{align}
where $(\cdot)_+$ denotes the projection onto the non-negative space, ensuring that each element of $\bm{\Lambda}$ is non-negative.

\subsubsection{Optimization of the PA positions $\mv V$} 
Finally, we optimize the PA positions $\mv V$ under given transmit power $\mv P$ and decoding vector $\mv w$, then problem (P1) is reformulated as 
\begin{subequations}\label{P4}
	\begin{align}
	\text{(P4)}: \mathop {\min }\limits_{\mv V} &   \|\mv{w}^{\rm H} \mv{G} (\mv{V})\mv P -{\mv 1}^{\rm T}\|^2 \nonumber \\
	\mathrm{s.t.}~& (11 {\rm b})~{\rm and}~(11 {\rm c}). \nonumber
	\end{align}
	\end{subequations}

To tackle problem (P4), we propose a Gauss-Seidel-based  iterative strategy, where the position of each PA is updated individually and sequentially while keeping the other PA positions fixed. Without loss of generality, we focus on optimizing the position of the $n$-th pinching element on waveguide $m$, denoted as $v_{m,n}$, while fixing all other positions $v_{m',n'}$ for $(m', n') \neq (m, n)$. The resulting optimization subproblem w.r.t. $v_{m,n}$ is formulated as
\begin{subequations}\label{P5}
	\begin{align}
	\text{(P5)}: \mathop {\min }\limits_{v_{m,n}} &   \|\mv{w}^{\rm H} \mv{G}_{m,n} (v_{m,n})\mv P -{\mv 1}^{\rm T}\|^2 \nonumber \\
	\mathrm{s.t.}~
	& v_{m,\tilde n} - v_{m,\tilde n-1} \ge L_0, \tilde n = n, n+1, \label{P5-b} \\
	& (11 {\rm b}), \nonumber
\end{align}
\end{subequations}
where $\mv G_{m,n}(v_{m,n})=[\mv g_1, \cdots, \mv g_{m-1}, \mv g_m(v_{m,n}), \mv g_{m+1}, \cdots, \\ \mv g_{M}]^{\rm T}$, with $\mv{g}_{m'}$, $\forall m' \neq m$, denoting the $m'$-th row of the channel matrix $\mv{G}(\mv{V})$. When the positions $v_{m',n'}$, $\forall (m', n') \neq (m, n)$, are fixed, the $m$-th row of $\mv{G}(\mv{V})$ becomes a function of $v_{m,n}$, denoted as $\mv{g}_m(v_{m,n})$, where its $k$-th element is 
\begin{align}
g_{m,k}(v_{m,n})= \alpha_{m,k}(v_{m,n})+\sum\nolimits_{n^{\prime}, n^{\prime} \neq n}\alpha_{m,k}(v_{m,n^{\prime}}),
\end{align}
where
\begin{align}
&\alpha_{m,k}(v_{m,n})=\nonumber \\
&\frac{\lambda e^{-j \left(\frac{2 \pi}{\lambda}\sqrt{(v_{m,n}-x_k)^2+((m-1)a-y_k)^2+d^2}+ \frac{2 \pi i_{ref}}{\lambda}v_{m,n}  \right)}}{4\pi \sqrt{(v_{m,n}-x_k)^2+((m-1)a-y_k)^2+d^2}}.
\end{align}

Next, we express the objective function of problem (P5) in terms of $\mv{g}_m(v_{m,n})$ and rewrite it as
\begin{align}
\mathop {\min }\limits_{v_{m,n}} \| ({\bar w}_{m} \mv g_m^{\rm T}(v_{m,n})+ \mv c) \mv P - {\mv 1}^{\rm T} \|^2,
\end{align}
where $\bar{ (\cdot)}$ is the conjugate operation, $\mv c= \sum_{m^{\prime},m^{\prime} \neq m} {\bar w}_{m^{\prime}} {\mv g}^{\rm T}_{m^{\prime}}$. Furthermore, by defining $\mv q= \mv c \mv P - \mv 1^{\rm T}$, and after algebraic manipulation, the objective function of problem (P5) becomes
\begin{align}
&\mathop {\min }\limits_{v_{m,n}} |w_m|^2 {\mv g}_m^{\rm T} (v_{m,n}){\mv P}^2 {\bar {\mv g}}_m (v_{m,n}) + 2{\rm Re} \big(  {\bar w}_m {\mv g}_m^{\rm T}(v_{m,n}) \mv P \mv q^{\rm H} \big) \nonumber \\
&= \mathop {\min }\limits_{v_{m,n}} \sum_k \bigg( |w_m|^2 p_k |g_{m,k}(v_{m,n})|^2 \nonumber \\
& \qquad \qquad + 2{\rm Re}\big( {\bar w}_m \sqrt{p_k} g_{m,k}(v_{m,n}) {\bar q}_k \big)    \bigg).
\end{align}

Then, by introducing $\beta_{m,k}=\sum_{n^{\prime}, n^{\prime} \neq n}\alpha_{m,k}(v_{m,n^{\prime}})$, problem (P5) is reformulated as 
\begin{subequations}\label{P6}
	\begin{align}
	\text{(P6)}: \mathop {\min }\limits_{v_{m,n}} &   \sum_k \bigg( A_k |\alpha_{m,k}(v_{m,n}) |^2+ 2{\rm Re}\big(B_k  \alpha_{m,k}(v_{m,n})\big) \bigg) \nonumber \\
	\mathrm{s.t.}~ & (11 {\rm b})~{\rm and}~(23 {\rm a}), \nonumber
\end{align}
\end{subequations}
where $A_k=|w_m|^2p_k$, $B_k= |w_m|^2 p_k {\bar \beta}_{m,k} +  {\bar w}_m\sqrt{p_k}{\bar q}_k$.

Problem (P6) is a scalar optimization problem over a closed interval, which can be solved by grid search efficiently. Specifically, we discretize the interval $[0,L_x]$ into $L$ equally spaced points and evaluate the objective function at each candidate value in the following set:
\begin{align}
{\cal G}_m \triangleq \left\{0,\frac{L_x}{L-1},\frac{2L_x}{L-1},\cdots,L_x \right\}, 
\end{align}
then the optimal $v^{*}_{m,n}$ is 
\begin{align}
&v^{*}_{m,n}= \mathop {\rm argmin }\limits_{v_{m,n} \in  {\hat{\cal G}}_m} \sum_k \bigg( A_k |\alpha_{m,k}(v_{m,n}) |^2 \nonumber \\
&\qquad \qquad+ 2{\rm Re}\big(B_k  \alpha_{m,k}(v_{m,n})\big) \bigg),
\end{align}
where ${\hat{\cal G}}_m$ is the feasible search region of $v_{m,n}$ satisfying \eqref{P1-b} and \eqref{P5-b}, i.e., 
\begin{align}
{\hat{\cal G}}_m={\cal G}_m \cap [v_{m,n-1}+L_0, L_x] \cap [0,v_{m,n+1}-L_0].
\end{align}
After obtaining the optimized PA positions $\mv{V}^*$, we iteratively update the decoding vector $\mv{w}$ and transmit power $\mv{P}$ until the decrease of the objective value falls below a predefined threshold. This iterative procedure yields a sequence of monotonically non-increasing objective values with corresponding solutions $\{ \mv{w}^*, \mv{P}^*, \mv{V}^* \}$, thereby ensuring the convergence of the proposed AO framework.

\begin{algorithm}[h]
\caption{Overall algorithm for solving problem (P1)}
\label{alg:AO}
\begin{algorithmic}[1]
\State \textbf{Initialize:} PA positions $v_{m,n}^{(0)} = \frac{L_x}{N+1} \times n$, transmit power $\mv{P}^{(0)}=\sqrt{P^{\rm max}}$, set iteration index $o = 1$.
\Repeat
    \State Optimize the decoding vector $\mv{w}^{(o)}$ by solving problem (P2) with $\mv{P}^{(o-1)}$ and $\mv{V}^{(o-1)}$.
    
    \State Optimize the transmit power $\mv{P}^{(o)}$ by solving problem (P3) with  $\mv{w}^{(o)}$ and $\mv{V}^{(o-1)}$.
    
    \For{each PA}
        \State Fix $v^{(o-1)}_{m',n'}$, $\forall (m', n') \neq (m, n)$
        \State Optimize $v_{m,n}^{(o)}$ by solving problem (P6).
    \EndFor

    \State $o=o + 1.$
\Until{the decrease of the objective value is below a threshold $\epsilon$.}
\State \textbf{Return:} Optimized $\mv{w}^*$, $\mv{P}^*$, and $\mv{V}^*$.
\end{algorithmic}
\end{algorithm}

	\begin{figure*}[h]
		\centering 
		\subfigure[]{\includegraphics[width=.24\textwidth]{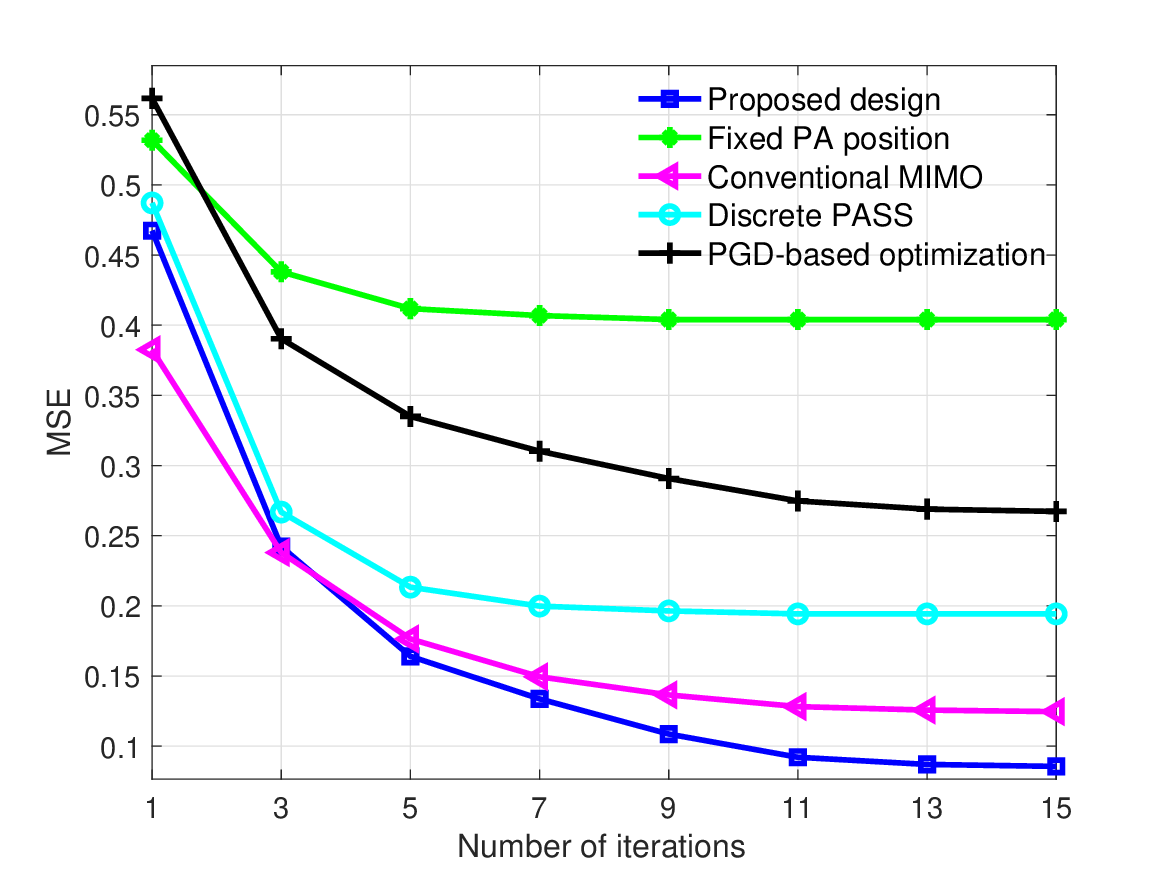}\label{fig:converge}}
			\subfigure[]{\includegraphics[width=.24\textwidth]{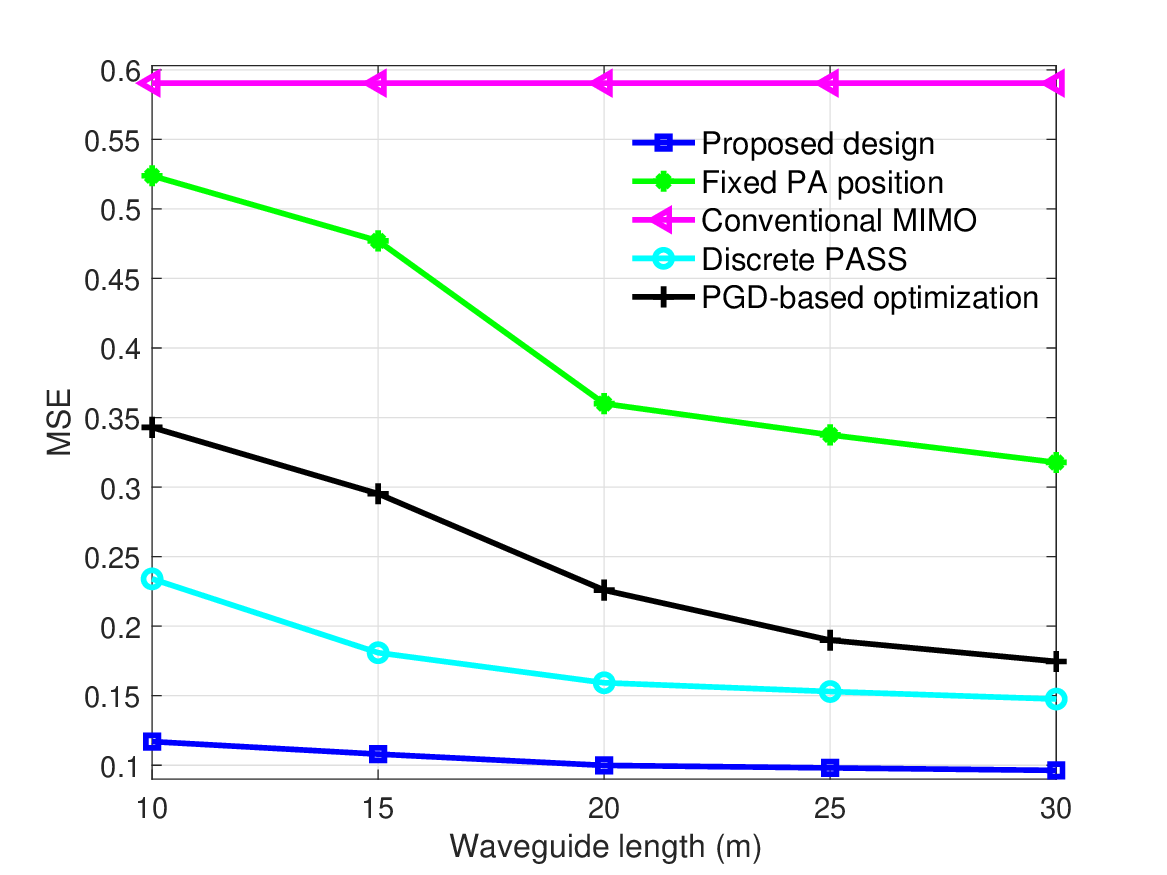}\label{fig:waveguidelength}}
			\subfigure[]{\includegraphics[width=.24\textwidth]{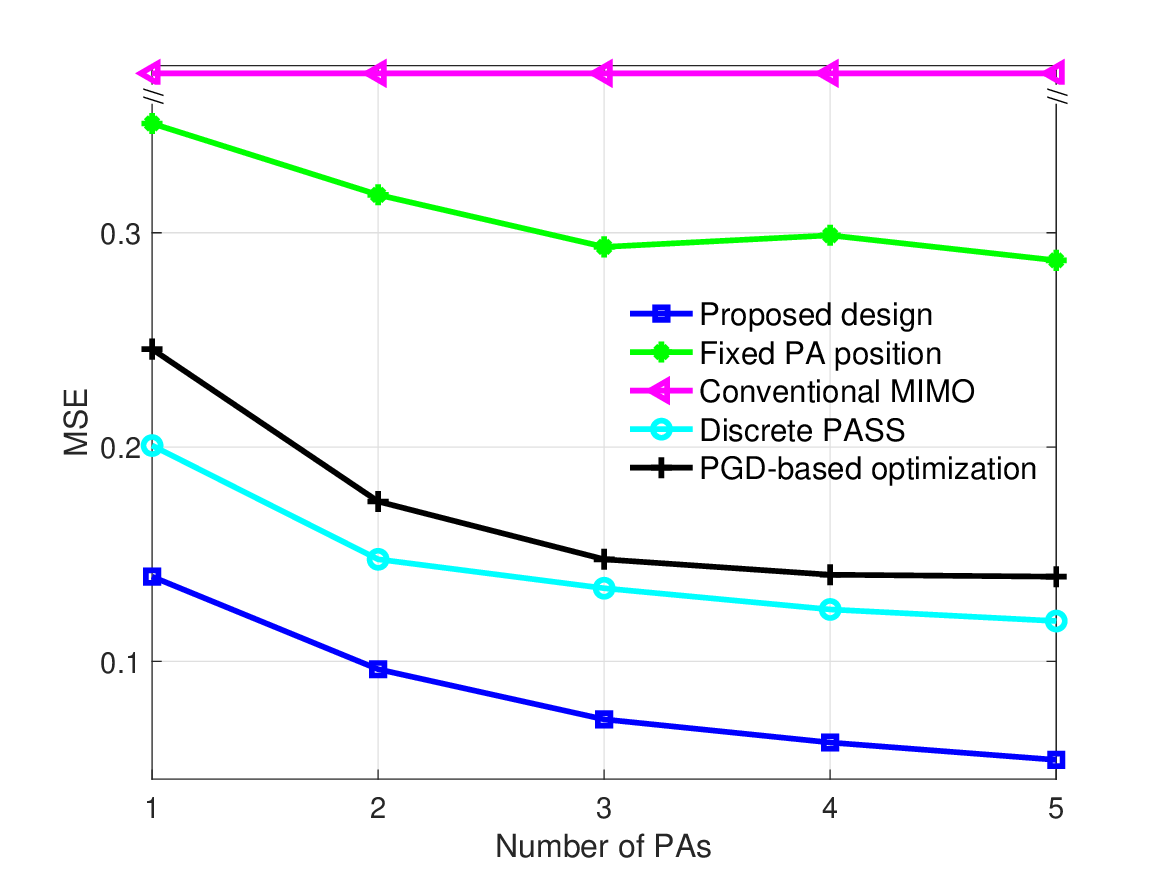}\label{fig:PAnum}}
			\subfigure[]{\includegraphics[width=.24\textwidth]{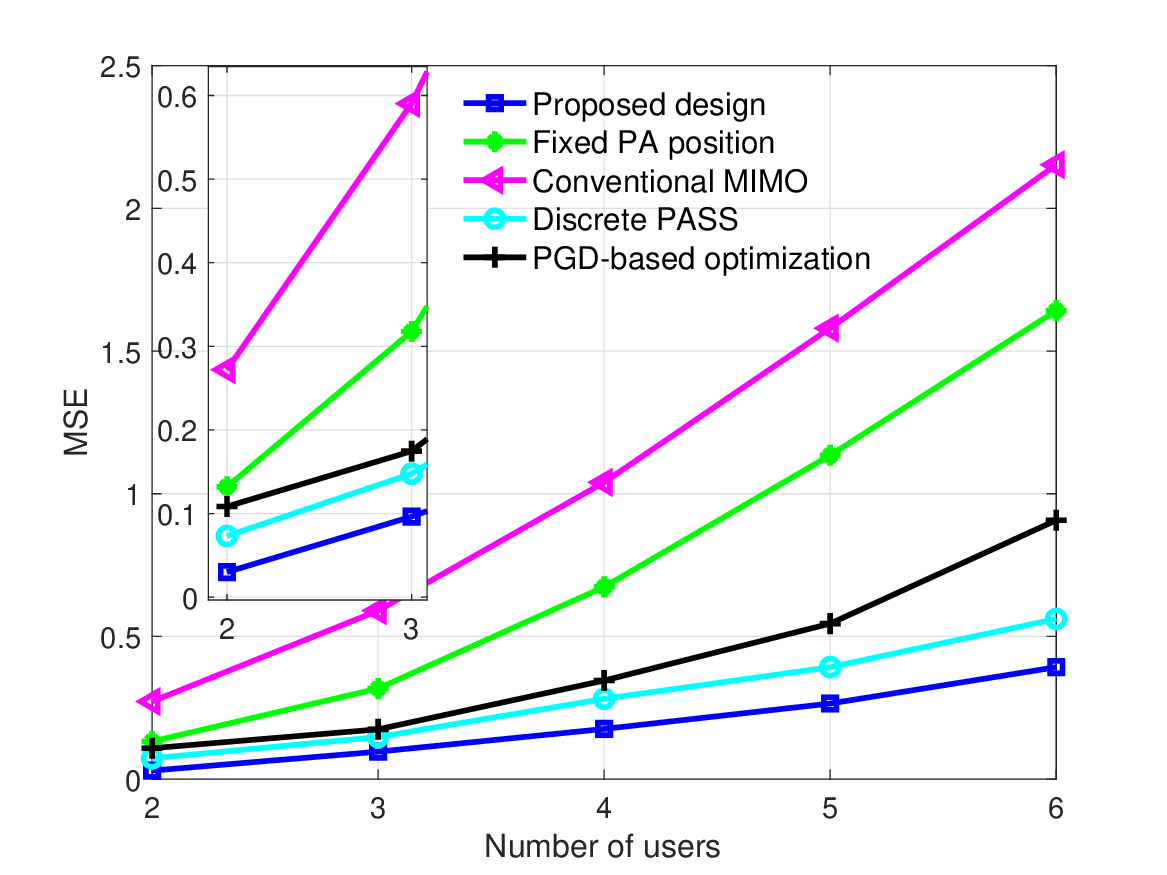}\label{fig:usernum}}
        \vspace{-5pt}
		\caption{(a) Convergence performance of our proposed design. (b) MSE w.r.t. the waveguide length. (c) MSE w.r.t. the number of PAs. (d) MSE w.r.t. the number of users.}\label{fig:performance}
		\vspace{-2mm}
	   \end{figure*}

\section{Numerical Results}
This section evaluates the performance of the proposed joint PA position and communication design for PASS-aided AirComp. We adopt a parameter setup similar to that in \cite{JZhao,ABereyhi}. Specifically, we consider a PASS system covering an area of $L_x = 20~\mathrm{m}$ and $L_y = 6~\mathrm{m}$, where $M = 4$ waveguides are deployed at a height of $d = 5~\mathrm{m}$, with an inter-waveguide spacing of $a = L_y / (M - 1)$. The carrier frequency is set as $f = 28~\mathrm{GHz}$. Each waveguide is equipped with $N = 2$ pinching elements and connected to a dedicated RF chain to serve $K = 3$ users. The minimum spacing between adjacent PAs is set to $L_0 = \lambda/2$, and the refractive index of the waveguide material is $i_{\mathrm{ref}} = 1.44$. The noise power is assumed to be $\sigma^2 = -90~\mathrm{dBm}$. Users are randomly distributed within the coverage area, and all numerical results are averaged over $300$ independent user location realizations. We consider the following benchmark schemes,
\begin{itemize}
    \item \textbf{Fixed PA position design:} PAs are uniformly distributed on each waveguide with fixed locations given by $v_{m,n} = \frac{L_x}{N+1} \times n$. Then we solve problem (P1) to optimize the transmit power $\mv{P}$ and decoding vector $\mv{w}$.
    \item \textbf{Conventional MIMO system:} We consider a standard MIMO system, where the BS is equipped with $M$ antennas, uniformly spaced at $\lambda/2$, each connected to a dedicated RF chain. For fairness, the number of antennas is set equal to the number of waveguides in the PASS system, as activating a PA on a waveguide incurs no additional cost.
    \item \textbf{Discrete PASS:} PAs are only allowed to be activated at preconfigured discrete locations. We consider 300 potential PA positions, uniformly spaced along each waveguide.
    \item \textbf{Projected Gradient Descent (PGD) based PA position optimization:} The PA positions in problem (P4) are optimized using PGD \cite{TVu}, while the transmit power $\mv{P}$ and decoding vector $\mv{w}$ are still obtained by solving problems (P2) and (P3), respectively.
\end{itemize}

First, Fig.~\ref{fig:converge} illustrates the convergence behavior of the proposed joint PA position and communication design. It is observed that the MSE achieved by our design decreases rapidly with the number of iterations and converges within approximately 15 iterations. Moreover, our proposed method consistently outperforms all benchmark schemes, demonstrating the effectiveness of the our joint optimization framework.

Next, Fig.~\ref{fig:waveguidelength} shows the MSE performance w.r.t. varying waveguide lengths $L_x$. It is observed from Fig.~\ref{fig:waveguidelength} that, as the waveguide length increases, the MSE decreases. This is because longer waveguides offer more design DoFs and wider coverage, allowing PAs to be flexibly positioned closer to the users. As a result, the system can achieve more favorable channels and reduce signal aggregation errors in AirComp. In contrast, conventional MIMO systems are unable to exploit these structural advantages, leading to worse performance. This highlights the benefit of PASS-enabled AirComp in extended service areas.

Then, Fig.~\ref{fig:PAnum} shows the MSE performance w.r.t the number of PAs $N$ per waveguide. It is observed that our proposed design significantly outperforms all benchmark schemes. Moreover, increasing $N$ leads to improved MSE performance, owing to the enhanced spatial DoFs enabled by pinching beamforming. In contrast, the fixed PA position design exhibits performance saturation, as inappropriate antenna placement can degrade the MSE. This further demonstrates that, with proper PA position optimization, PASS-enabled AirComp can leverage more favorable channels for efficient data aggregation.

Finally, Fig.~\ref{fig:usernum} shows the MSE performance under different numbers of users $K$. The MSE increases with $K$ due to the growing complexity of signal alignment in AirComp. Nevertheless, our proposed design outperforms all benchmarks, with increasing performance gaps as $K$ grows.

\section{Conclusions}
This letter has introduced a novel framework for AirComp, i.e., PASS-aided AirComp, where spatially reconfigurable PAs are leveraged to enhance channel conditions for accurate data aggregation. We have jointly optimized the PA position, transmit power, and decoding vector to minimize the AirComp MSE, which is highly non-convex and non-trivial to solve in general. To tackle the problem, we have proposed an AO framework, where PA positions are iteratively refined via a Gauss-Seidel strategy. Simulation results have verified the superiority of the proposed design over various benchmark schemes.

\end{document}